\documentclass[12pt,amsart]{iopart}
\usepackage{fullpage,graphicx,epsf}

\usepackage{iopams}%
\usepackage{amssymb}
\usepackage{amscd}
\usepackage{color}%
\usepackage{graphicx}
\usepackage{epsfig}
\usepackage[english]{babel}
\usepackage{amsfonts}

\begin{document}
\title{Spontaneous symmetry breaking in a bridge model fed by junctions  }
\author{Vladislav Popkov$^{1,2}$, Martin R. Evans $^{3}$ and David Mukamel$^{4}$}
\address{ $^1$ Dipartimento di Fisica "E.R. Caianiello", and
Consorzio Nazionale Interuniversitario per le Scienze Fisiche
della Materia (CNISM), Universit\`a di Salerno, Baronissi,
Italy\\
$^2$ Interdisziplin\"ares Zentrum fur Komplexe Systeme,
R\"omerstrasse 164, D-53117 Bonn, Germany\\
$^3$ SUPA, School of Physics, University of Edinburgh, Mayfield Road, Edinburgh EH9 3JZ, UK\\
$^4$ Department of Physics of Complex Systems, The Weizmann
Institute of Science, Rehovot 76100, Israel }
\ead{popkov@sa.infn.it,m.evans@ed.ac.uk,david.mukamel@weizmann.ac.il}

\today

\begin{abstract}
We introduce a class of 1D models mimicking a single-lane bridge
with two junctions and two particle species driven in opposite
directions. The model exhibits spontaneous symmetry breaking (SSB)
for a range of injection/extraction rates. In this phase the steady
state currents of the two species are not equal. Moreover there is
a co-existence region in which the symmetry broken phase co-exists
with a symmetric phase. Along a path in which the extraction rate is
varied, keeping the injection rate fixed and large, hysteresis takes
place.  The mean field phase diagram is calculated and supporting
Monte-Carlo simulations are presented. One of the transition lines
exhibits a kink, a feature which cannot exist in transition lines of
equilibrium phase transitions.

\end{abstract}

\maketitle

\section{Introduction and Model Definition}
Nonequilibrium stationary states (NESS), in which probability currents
are supported and invariant measures are not generally of
Gibbs-Boltzmann form, offer many surprising phenomena not seen in
equilibrium systems.  Examples include boundary-induced phase
transitions and spontaneous symmetry breaking (SSB) in one-dimensional
systems.  The minimal models for NESS are driven particle models such
as the Totally Asymmetric Exclusion Process (TASEP) and its relatives.
Some exact solutions for these models have been found, for example the
TASEP, leading to an understanding of boundary-induced phase
transitions \cite{BE07}.
However, important questions remain as to the nature of
SSB in driven one-dimensional systems.

The SSB phenomenon was first observed  in what is often referred to
as the bridge model \cite{EFGMPRL}, and subsequently in some other
models \cite{PP01,PK07}.  The bridge model comprises a one
dimensional lattice on which positive particles (pluses) move to the
right and negative particles (minuses) move to the left. At the left
boundary of the lattice, pluses may enter the lattice and minuses
may leave; at the right boundary of the lattice minuses may enter
the lattice and pluses may leave. When the exit rate at which
particles may leave the lattice  is lowered, the stationary state
changes from a symmetric one, in which the currents and bulk
densities of pluses and minuses are equal, to a symmetry-broken
state where there is a majority species of particle with larger
current and higher bulk density than the minority species.  However,
the symmetry broken state has not been solved exactly except in the
limit in which  the exit rate tends to zero where it has been
rigorously proven that SSB occurs\cite{GLEMSS95}.

The transition to this symmetry broken state remains a subject of
debate.  A mean field theory originally predicted that the
transition should occur via an intermediate, weakly symmetry-broken
phase \cite{EFGMJSP}. Monte Carlo simulations have shown that the
mean field theory does not correctly predict the position of the
transition \cite{AHR98}, however, at least on finite systems, an
intermediate phase is seen and the transition from symmetric to the
strongly symmetry-broken phase occurs through a  sequence of
transitions \cite{CEM01}.  This sequence has also been observed in a
related model \cite{PK07}.  However, it has been suggested that the
region of parameter space occupied by the intermediate phase and
over which the sequence of transitions occurs disappears in the
infinite system size limit \cite{EPSZ05}.

The failure of the mean-field approximation
to exactly predict the phase diagram can
be traced back to the boundary conditions of the bridge model which do
not correspond to particle reservoirs at fixed density.  Instead there
are effective impurities at the boundaries \cite{PS04}.
Also the hydrodynamic limit of the bridge model
is not well-defined over  the whole phase space.
In particular, it breaks down at the SSB
transition in the   limit of small input and exit rates\cite{PS04}.

In this paper we introduce a new class of bridge models
demonstrating SSB phenomenon.  In these models the input and output
of the pluses and minuses are governed by TASEPs. Thus one can think
of the ends of the bridge as junctions where TASEPs for the pluses
and minuses merge. The input and output rates at the bridge are not
external parameters as in TASEP, rather they are determined self
consistently by the dynamics of the bridge and its feeding segments.
The models exhibit complex phase diagrams  including a phase
co-existence region showing hysteresis and quite special triple
phase co-existence point, which appear to be correctly predicted by
a mean field theory.

\begin{figure}
[ptb]
\begin{center}
\includegraphics[
height=4.5cm
]%
{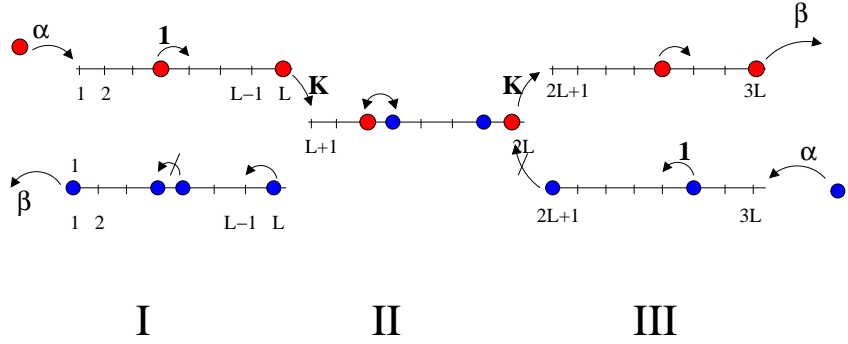}%
\caption{The bridge model with two junctions. Positively
(negatively) charged particles hop to the right (left). The model is
invariant with respect to left-right reflection and charge
inversion. Section II is the bridge. It contains positive and
negative particles  and holes. Sections I and III comprise parallel
segments
each containing pluses and holes or minuses and holes.}%
\label{Fig_SSBmodel}%
\end{center}
\end{figure}

Our model is defined as follows. We consider a chain of the length
$L$ occupied by two species of particles, ``plus" particles moving
to the right and ``minus" particles moving to the left, with
hardcore exclusion and random sequential update. The chain is region
II of Fig.~\ref{Fig_SSBmodel} and we shall refer to it as the
bridge. At each end of the bridge, there are junctions where the
chain splits into two parallel segments, one containing only plus
particles and holes and the other only minus particles and holes
(sections I and III in Fig.\ref{Fig_SSBmodel}). At the {\em
external} boundaries of the parallel segments, usual TASEP rules are
applied.  Namely, plus particles are injected into their segment of
section I at the left with rate $\alpha$, if the first site is
empty, and are removed with rate $\beta$ from the right end (upper
segment of section III in Fig.~\ref{Fig_SSBmodel}), if the last site
is occupied.  Likewise minus particles are injected into their
segment of section III with rate $\alpha$ and removed with rate
$\beta$ from the left end of their segment of section I.  In
sections I, III pluses hop to the right inside their segments with
rate 1. Plus particles enter the bridge at the left junction if the
first site of section II is empty and leave it at the other junction
if the first site of the plus segment of section III is empty, both
with rate $K$. Inside the bridge, particles exchange with empty
sites and with the minuses with the same rate $1$. Similarly, inside
the lattice minus particles hop to the left with rate 1 in sections
I and III, enter and leave the bridge if the entrance or exit site
is empty  with rate $K$, and inside the bridge exchange with both
pluses and empty sites with rate $1$. The model is symmetric with
respect to simultaneous charge inversion and left-right reflection.

In this work we
consider only $K=1$, leaving $\alpha$ and $\beta$ as the model
parameters. We calculate the phase diagram of the model using a mean
field approximation and direct simulation of the dynamics.

\begin{figure}
[ptb]
\begin{center}
\includegraphics[
height= 6.8cm
]%
{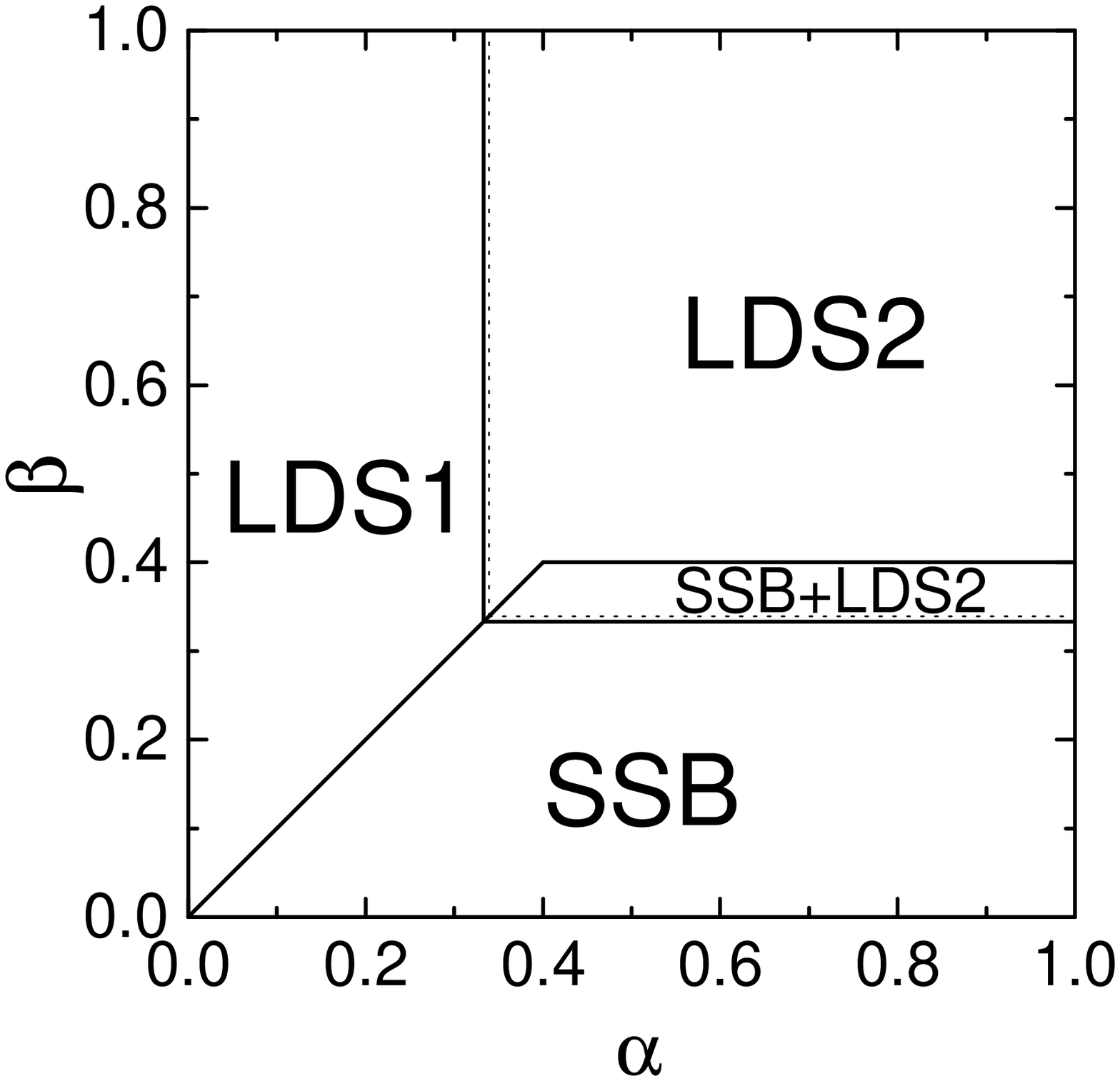} \caption{Mean field phase diagram of the
bridge model fed by  junctions for $K=1$ (see text for the
definition of the phases). The dotted lines show the location of
the LDS1-LDS2
and LDS2 $\rightarrow$ SSB phase transitions from the Monte-Carlo simulations. }%
\label{Fig_SSBphasediagram}%
\end{center}
\end{figure}

\section{Phase Diagram and Phase Transition Lines}
Before discussing the phase diagram of our model, it is useful to
recall the phase diagram of the TASEP (for plus particles moving to
the right). The phases are distinguished by the large system size
limits of the expressions for $\rho$ the bulk density of particles
far from the boundaries and $j= \rho(1-\rho)$ the current of
particles. When $\alpha < 1/2$ and $\beta>\alpha$ the low density
(LD) phase occurs where the density in the bulk is equal to $\alpha$
and is determined by the left boundary; when $\beta < 1/2$ and
$\alpha > \beta$ the high  density (HD) phase occurs where the
density in the bulk is equal to $1-\beta$ and is determined by the
right boundary; when $\alpha >1/2$ and $\beta >1/2$ the maximal
current (MC) phase occurs where the bulk density is $1/2$.

We now present the phase diagram of the bridge model fed by junctions
obtained from a  mean field
analysis detailed  in the following section. The resulting
$\alpha$--$\beta$ phase diagram is given in
Fig.~\ref{Fig_SSBphasediagram}; it contains three different phases.

\begin{figure}
[ptb]
\begin{center}
\includegraphics[
height=8.0cm
]%
{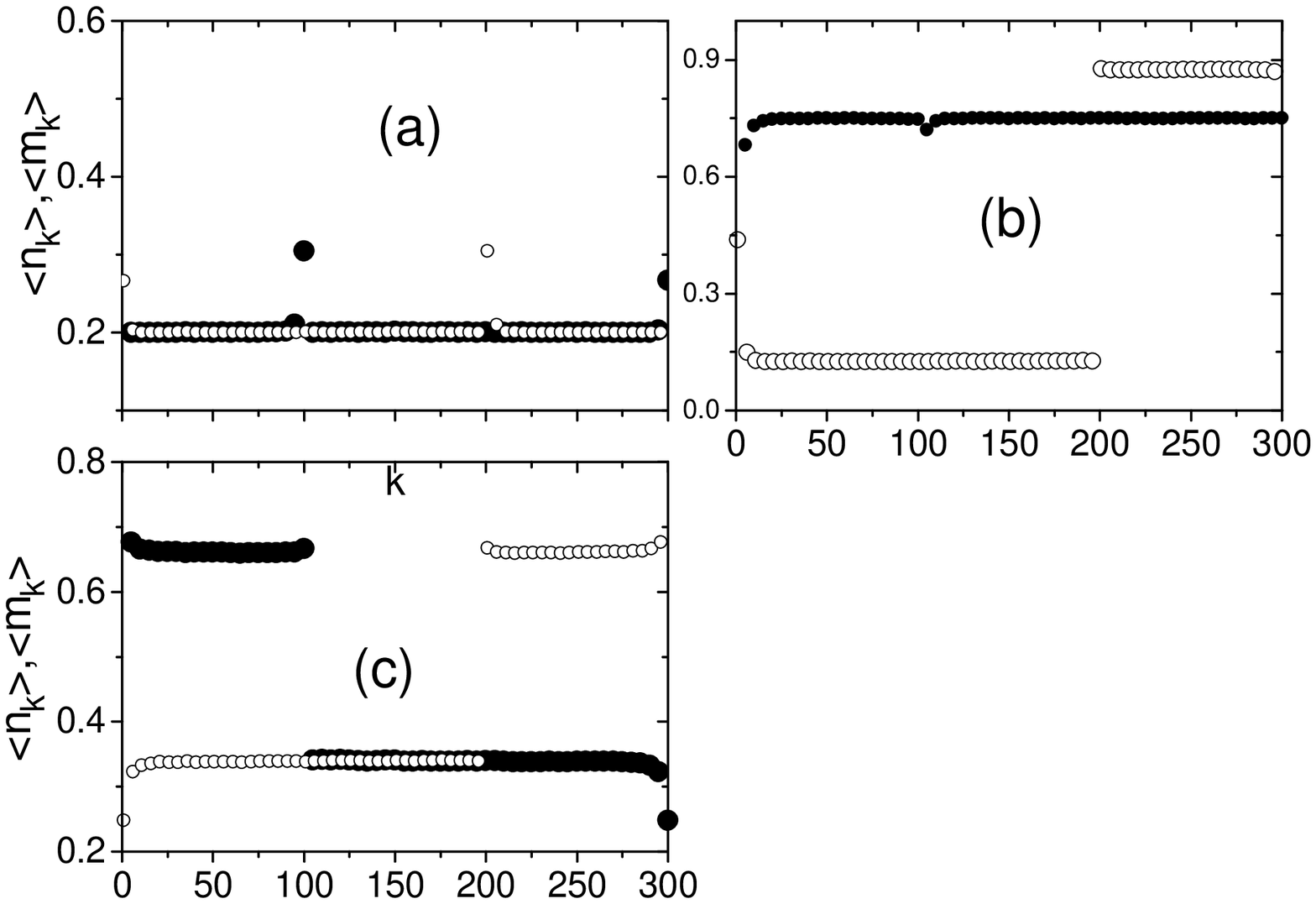}%
\caption{ Average density profiles for pluses and minuses, from
Monte-Carlo simulations, in the LDS1 phase (Panel (a)), in the SSB
phase with pluses majority (Panel (b)) and in the LDS2 phase (Panel
(c)). Pluses (minuses) correspond to closed (open) circles.  The
system of $300$ sites was equilibrated and then, averaging over
$10^{6}$ Monte-Carlo steps was done.  Parameters:(a)
$\alpha=0.2,\beta=0.4$; (b) $\alpha=0.2,\beta=0.25$; (c)
$a=0.8,\beta=0.9$.
}%
\label{Fig_AllPhases4}%
\end{center}
\end{figure}

\begin{description}
\item[(a) Low density symmetric (LDS1) phase] $\left(
\alpha<1/3,\beta>\alpha\right)$. Each species establishes a
homogeneous state with low particle density $\rho =\alpha$ in each
segment, see Fig.\ref{Fig_AllPhases4}(a).

\item[(b) Spontaneous symmetry-broken (SSB) phase] $\left(\beta<1/3,\alpha
>\beta\right)  $. The two species have different densities and fluxes
and the phase comprises  two symmetry related states. The majority
species  (the pluses in Fig.\ref{Fig_AllPhases4}(b)) establishes a
high density state with bulk density $1-\beta$ in all segments
while the minority species (the minuses in
Fig.\ref{Fig_AllPhases4}(b)) has bulk  density $\beta/2$ in the
bridge and section I, and bulk density $1-\beta/2$ in section III.

\item[(c) Low density symmetric (LDS2) phase]
$\left( \alpha>1/3,\beta>1/3\right)$. The
pluses have bulk density 2/3 in section I and bulk density 1/3 in
sections II,III whereas the minuses have bulk density 1/3 in
 sections I,II and 2/3 in section III.  Thus the profile of the
 minuses mirrors that of the pluses, see Fig.\ref{Fig_AllPhases4}(c), and the
 phase is symmetric. Note that while both densities on the bridge
 (section II) are low, on the other sections I,III
the  density of one of the species is low while that of the other
species is high. This phase is thus different from the LDS1 phase.
\end{description}

The phase diagram (Fig.~\ref{Fig_SSBphasediagram}) is reminiscent of
the TASEP phase diagram discussed above, but with several important
differences. First the LDS2 phase replaces the maximal current phase
and the transition lines to the LDS2 phase are at $\alpha=1/3$,
$\beta=1/3$ rather than $\alpha=1/2$, $\beta=1/2$.  Secondly, the
SSB phase replaces the high density phase of the TASEP. Clearly a
high density symmetric phase is not possible in the present model
since in that phase the densities of both species on the bridge
section have to be greater than $1/2$. Thirdly, there is a
co-existence  region $\left(1/3<\beta<2/5,\alpha>\beta\right)$ where
the system may be in either of the LDS2 and SSB phases; this produces
an interesting kink in the phase boundary of the SSB phase.

We now discuss the phase transitions and transition lines separating
the phases described above. For the TASEP, as well as other
nonequilibrium steady states with a non-zero conserved current, the
`order' of a phase transition is determined by the non-analyticity
of the current at the transition, i.e. the order of the derivative
of the current at which a singularity appears \cite{BE07}.
In the present model we find some novel features for the
nonequilibrium phase transitions. Notably, across {\em all} the
transition lines some of the bulk densities change discontinuously,
as can be seen by comparing stationary densities in different
phases. Indeed for the LDS1/SSB transition there is even a
discontinuity in the current of the
minority species. We now describe the behaviours at the transitions. \\

\noindent{\bf LDS1/LDS2 transition line}: $\alpha=1/3,\; \beta
>\alpha$. The  bulk densities in the bridge are continuous across
the transition and the currents are continuous. In section I the
plus density jumps discontinuously across the transition from 1/3 to
2/3, whereas the  minus density is continuous. Similarly in section
III, the minus density jumps discontinuously from 1/3 to 2/3. On the
transition line one finds a shock in section I separating regions of
plus densities 1/3 and 2/3 and a shock in section III separating
regions of minus densities 2/3 and 1/3.\\

\noindent{\bf LDS1/SSB transition line}: $\alpha=\beta < 1/3$.  The
bulk densities in all sections are discontinuous across the
transition: the bulk density of the majority species jumps from
$\alpha$ to $1-\alpha$ in all sections; the bulk density of the
minority species (taken as minus) jumps from $\alpha$ to $\alpha/2$
in sections I and II and from $\alpha$ to $1-\alpha/2$ in section
III. The majority current is continuous whereas the minority current
jumps from $\alpha(1-\alpha)$ to $\alpha/2(1-\alpha/2)$. On the
transition line two  types of shock configurations are in fact
observed. First there is  symmetric
configuration where the plus density is $\alpha$ in sections I and
II, and in section III there is a shock separating regions of plus
densities $\alpha$ and $1-\alpha$. Similarly the minus density is
$\alpha$ in sections II and III, and in section I there is a shock
separating regions of minus densities $1-\alpha$ and $\alpha$.
Secondly, there is an asymmetric shock configuration where, taking
the majority species to be plus, the bulk plus density is $1-\alpha$
in sections II and III and in section I there is a shock between
regions of plus density $\alpha$ and $1-\alpha$. The minority
species density, taken to be minus, is   $\alpha/2$ in sections I
and II and $1-\alpha/2 $ in section III.
Here the currents of plus and minus are unequal.\\

\noindent{\bf LDS2/SSB Co-Existence Region}: $\alpha>\beta$,
$1/3\leq\beta\leq 2/5\,$. In this region, denoted "SSB+LDS2" in the
phase diagram Fig.\ref{Fig_SSBphasediagram}, the system can be in
either of the two symmetry-related SSB states or the LDS2 phase. In
the stochastic model of infinite size, the three phases are stable.
In finite systems flips between the phases take place,
with typical time between the flips growing exponentially with the system size
$L$ (this issue will be addressed in Sec.\ref{timescale}). Note that
spatial co-existence of the phases (like e.g. in \cite{PP01}) is not
possible since they carry different currents.

\begin{figure}
[ptb]
\begin{center}
\includegraphics[
height=5.8cm
]%
{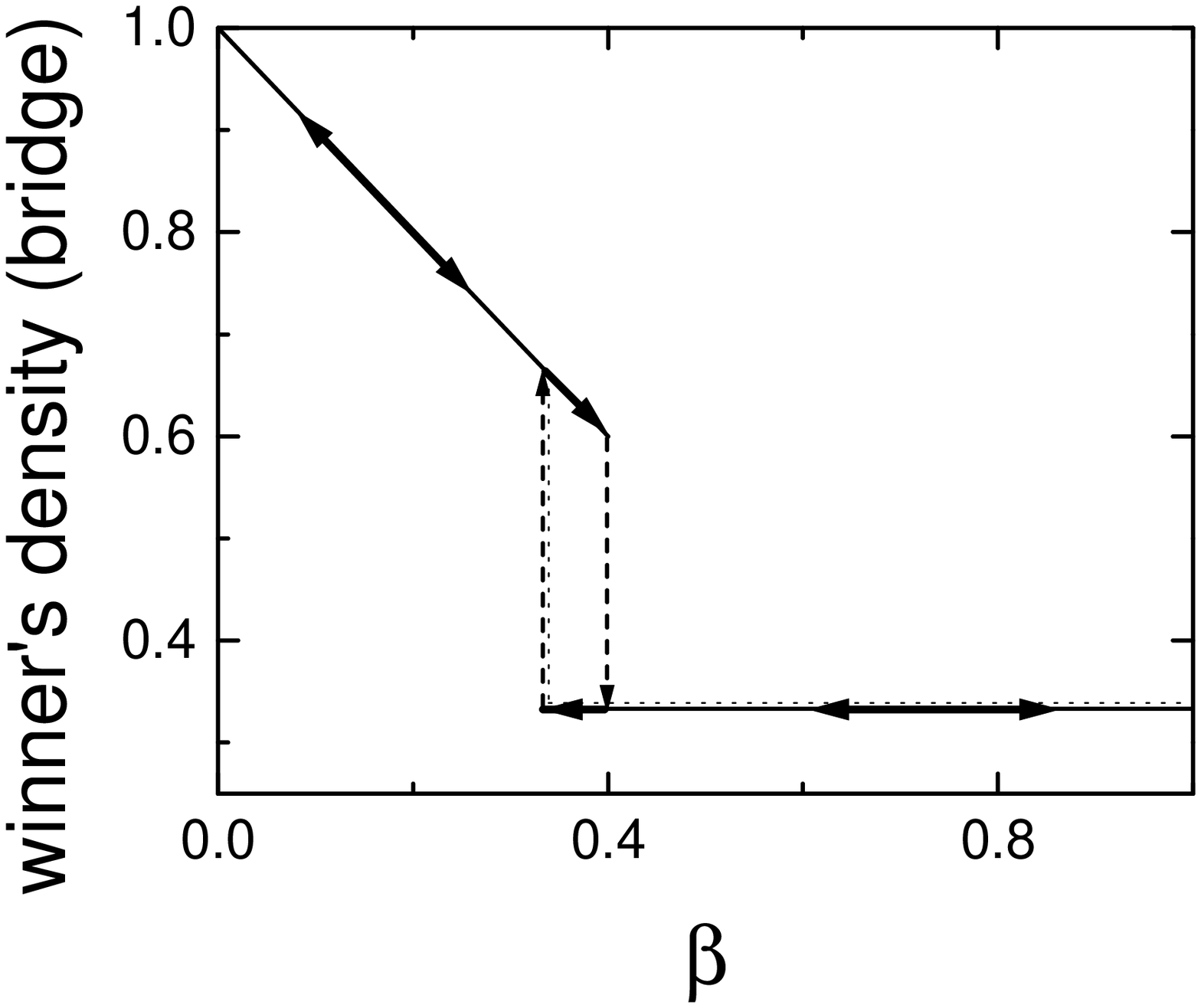}%
\caption{Hysteresis path for the majority  density in the bridge
section in the coexistence region: keeping the injection rate
constant $\alpha>2/5$, the extraction rate $\beta$ was changed from
$0$ to $1$ and back. The majority density in the bridge segment is
shown as function of $\beta$. On increasing $\beta$, SSB phase is
stable for $\beta<2/5$, while on the way back (decreasing $\beta $)
the LDS2 phase ($p=m=1/3$) is stable for $1/3<\beta<1$.
The (barely visible) dotted line  is the result of Monte-Carlo simulations}%
\label{Fig_SSBhysteresis}%
\end{center}
\end{figure}

In the infinite system, the co-existence region SSB+LDS2 entails
hysteresis, which is observed when keeping the injection rate
$\alpha>2/5$ constant and increasing/decreasing adiabatically the
extraction rate $\beta$ from $0$ (SSB phase) to $1$ (LDS2)  and
back. Indeed, along the path $0\rightarrow1$ the SSB/LDS2 transition
happens at $\beta=2/5$, while on the backward path it will happen at
$\beta=1/3$. The hysteresis is illustrated in
Fig.\ref{Fig_SSBhysteresis}.

A kink point with coordinates  $\alpha=\beta=2/5$ on the
phase diagram is an unusual one. In phase diagrams of systems at
equilibrium, such a kink would imply that there must be another
first order transition line emerging from that point, making it a
triple point. This is a result of the Clausius-Clapeyron relation.
Our model is out of equilibrium, and the existence of a kink,
although unusual, does not violate any rule.

\section{Mean field solution}
We now derive the phase diagram  through a mean field
approximation commonly used for one-dimensional stochastic
systems, wherein two-point correlation functions are replaced by
products of one-point correlation functions.  For the TASEP this
mean-field approximation is known to predict the correct phase diagram
and stationary bulk densities \cite{DDM92}.

Let us denote the stationary density of plus
particles at site $k$ as $p_{k}$. Then, the mean field approximation
gives the stationary flux of
pluses  as
\begin{equation}
j_{+} = p_{k}(1-p_{k+1})=p_{S}(1-p_{S}) \label{j+}
\end{equation}
where  $S$=I,II,III and $p_{S}$ is the limiting bulk density far
away from the boundaries in segment $S$.
Similarly,
the  density of minus
particles at site $k$ is $m_{k}$
and  we write
\begin{equation}
j_{-}=  (1-m_{k})m_{k+1}=m_{S}(1-m_{S})\;.
\label{j-}
\end{equation}
In the stationary state the bulk densities $p_{S},m_{S}$ can
be different  in  different segments $S$ but the
currents $j_{+}$,$j_{-}$  have the same values everywhere.
At the boundaries the currents read
\begin{eqnarray}
j_{+}  &  =\alpha(1-p_{1})=\beta p_{3L}\label{B+}\\
j_{-}  &  =\alpha(1-m_{3L})=\beta m_{1} \label{B-}%
\end{eqnarray}
and at the junctions,%
\begin{eqnarray}
j_{+}  &  =p_{L}(1-p_{L+1}-m_{L+1})=p_{2L}(1-p_{2L+1})\label{J+}\\
j_{-}  &  =m_{2L+1}(1-p_{2L}-m_{2L})=m_{L+1}(1-m_{L})\;. \label{J-}%
\end{eqnarray}
Following  \cite{EFGMJSP,PK05}, we define effective entrance rates $\alpha_{S}^{+}%
,\alpha_{S}^{-}$ and exit rates $\beta_{S}^{+},\beta_{S}^{-}$ for
each  segment. For plus particles
\begin{eqnarray}
\alpha_{I}^{+}  &  =\alpha\;;\mbox{ \ \ }\alpha_{II}%
^{+}=\frac{j_{+}}{1-p_{L+1}}\;;\mbox{ \ \ }\alpha_{III}^{+}=\frac{j_{+}%
}{1-p_{2L+1}}=p_{2L}\;;\mbox{\ }\label{al+}\\
\beta_{I}^{+}  &  =\frac{j_{+}}{p_{L}}=1-p_{L+1}-m_{L+1}\;;\mbox{ \ \ \ \ \ }\beta_{II}^{+}%
=\frac{j_{+}}{p_{2L}}= 1-p_{2L+1}\;;\mbox{ \ \ \  }\beta_{III}^{+}
=\beta. \label{be+}%
\end{eqnarray}
The rates $\alpha_{S}^{-},\beta_{S}^{-}$ for minus particles are
obtained by the substitutions $I\rightleftarrows III$ and
$p_{k}\rightarrow m_{3L-k+1}$.
For each species, each segment $S$ can be viewed as a TASEP model
with, e.g. for the pluses, the effective injection/extraction rates
$\alpha_{S}^{+}$, $\beta_{S}^{+}$. In the large segment length
limit, the bulk density of pluses $p_{S}$ in  segment $S$ can be
read from the phase diagram of a TASEP \cite{Schu93,Derr93}, namely
\begin{eqnarray}
\mbox{LD phase}\quad&\alpha_{S}^{+} <\beta_{S}^{+}\;,\; \alpha_{S}^{+}<
1/2
& \quad p_{S}=\alpha_{S}^{+}\;,\label{ASEP_LD}\\
\mbox{HD phase}\quad&\alpha_{S}^{+}   >\beta_{S}^{+}\;,\;\beta_{S}^{+}<
1/2 &\quad p_{S}=1-\beta_{S}^{+}
\label{ASEP_HD}\;,\\
\mbox{MC phase}\quad
&\alpha_{S}^{+},\beta_{S}^{+}    > 1/2&\quad
p_{S}= 1/2 \label{ASEP_MC}\;.
\end{eqnarray}
In the HD phase the density profile is flat at the right boundary
and in the LD  phase the density profile is flat at the left
boundary. For minuses, the corresponding bulk densities are given by
(\ref{ASEP_LD}--\ref{ASEP_MC}) with substitutions $+\rightarrow-$
and $p\rightarrow m$. Thus, in the large segment length limit, each
segment of the bridge model must exhibit  one of the solutions
(\ref{ASEP_LD}--\ref{ASEP_MC}). The possible solutions of the
mean-field equations \ref{j+}--\ref{J-} (MFE) are listed below.

(a) LDS1 phase.
Each segment is in the low density phase (\ref{ASEP_LD}) with density
$p_S = m_S=\alpha$ and currents $j_{+}= j_{-}=\alpha(1-\alpha)$.
The LD density profile implies that $p_{L+1}=p_{2L+1}=m_{2L}=m_L=\alpha$
and the symmetry of the phase implies that $m_{L+1}=p_{2L}$, $m_{2L+1}=p_L$.
As the solution is symmetric we need only  consider the effective rates
$\alpha_S^+$, $\beta_S^+$:
we read off from (\ref{al+}) $\alpha_{II}^+=\alpha_{III}^+=\alpha$
and from (\ref{be+}) $\beta_I^+=1-p_{L+1}-m_{L+1} = 1-2\alpha$,
$\beta_{II}^+=1-\alpha$. Conditions  (\ref{ASEP_LD}) for each segment
reduce to the key conditions
\begin{eqnarray}
\alpha &  <\beta,\mbox{ \ \ }\alpha<1/3
\end{eqnarray}
The first inequality follows from $\alpha_{III}^+ < \beta_{III}^+$
and the second follows from $\alpha_I^+ < \beta_I^+$.

(b) SSB phase.
Let us consider the solution of the MFE where
the pluses establish a high
density phase $p^+_S = 1-\beta> 1/2$ in all segments.
The minuses (the minority species), are in a low density
phase $m_{S}=\gamma$
in segments $S=I,II$ and
a high density  $m_{III}=1-\gamma$ in  segment III
giving  a  current $j_{-}=\gamma\left( 1-\gamma\right)  <j_{+}$.
The value of $\gamma$ is to be determined.
The structure of the profiles imply that $p_L = p_{2L}= 1-\beta$,
$m_L = m_{2L}= \gamma$, $m_{2L+1} = 1- \gamma$.
Then (\ref{J+},\ref{J-}) imply that $p_{2L+1}=1-\beta$,
$m_{L+1}=\gamma$, $p_{L+1}=1-\beta-\gamma$.
>From (\ref{J-}) one has
$j_{-}=\gamma\left(  1-\gamma\right) =\left( 1-\gamma\right) \left(
1-\left(  1-\beta\right) -\gamma\right)  $ giving  $\gamma=\beta/2$.
We read off from (\ref{al+}) $\alpha_{II}^+= 2(1-\beta)/3$,
$\alpha_{III}^+=1-\beta$
and from (\ref{be+}) $\beta_I^+=\beta_{II}^+=\beta$.
Similarly we find
$\alpha_{I}^-= \alpha_{II}^-=\beta/2$
and $\beta_{II}^- =1-\beta/2$, $\beta_{III}^- =\beta/2$.
Conditions  (\ref{ASEP_LD}) for each segment
reduce to the key conditions
\begin{eqnarray}
\alpha &  >\beta,\mbox{ \ \ }\beta<2/5\;,
\label{SSBstability_region}
\end{eqnarray}
the first coming from  $\beta_I^+ <\alpha_I^+$
and the second from  $\beta_{II}^+ <\alpha_{II}^+$.
By a symmetry
transformation (spatial inversion  and interchange of minuses and pluses) one
obtains the other symmetry-broken solution with
minuses in the majority.

(c) LDS2  phase.
In this solution  each species is in
a low density phase in the bridge and the segment fed by  the bridge and is
in a high density phase in the segment feeding  the bridge.
 Thus $p_I=1-\delta$,
$m_{III}=1-\delta$, $p_S=m_S=\delta$ (remaining segments) where
the value of $\delta<1/2$ is to be determined.
As this solution is symmetric we need only  consider the effective rates
$\alpha_S^+$, $\beta_S^+$.
The structure of the profiles implies that $p_L =1-\delta$,
$p_{L+1}=p_{2L+1} = \delta$,
$m_L = m_{2L}= \delta$, $m_{2L+1} = 1- \delta$.
Then (\ref{J+}) gives $j^+ = (1-\delta)(1-\delta-p_{2L})=p_{2L}(1-\delta)$,
implying that $p_{2L}= \delta$ and $\delta=1/3$. We
read off from (\ref{al+})
$\alpha_{II}^+= \alpha_{III}^+=\delta$
and from (\ref{be+}) $\beta_I^+=\delta$,
$\beta_{II}^+=1-\delta$.
Conditions  (\ref{ASEP_LD}) for each segment
reduce to the key conditions
\begin{eqnarray}
\alpha> 1/3\;,\; \beta  >1/3 
\label{HLSstability_region}
\end{eqnarray}%
the first coming from  $\beta_I^+ <\alpha_I^+$
and the second from  $\alpha_{III}^+ <\beta_{III}^+$.

\begin{figure}[ptb]
\centerline{
\includegraphics[width=6.3cm,height=4.3cm,angle=0,clip]{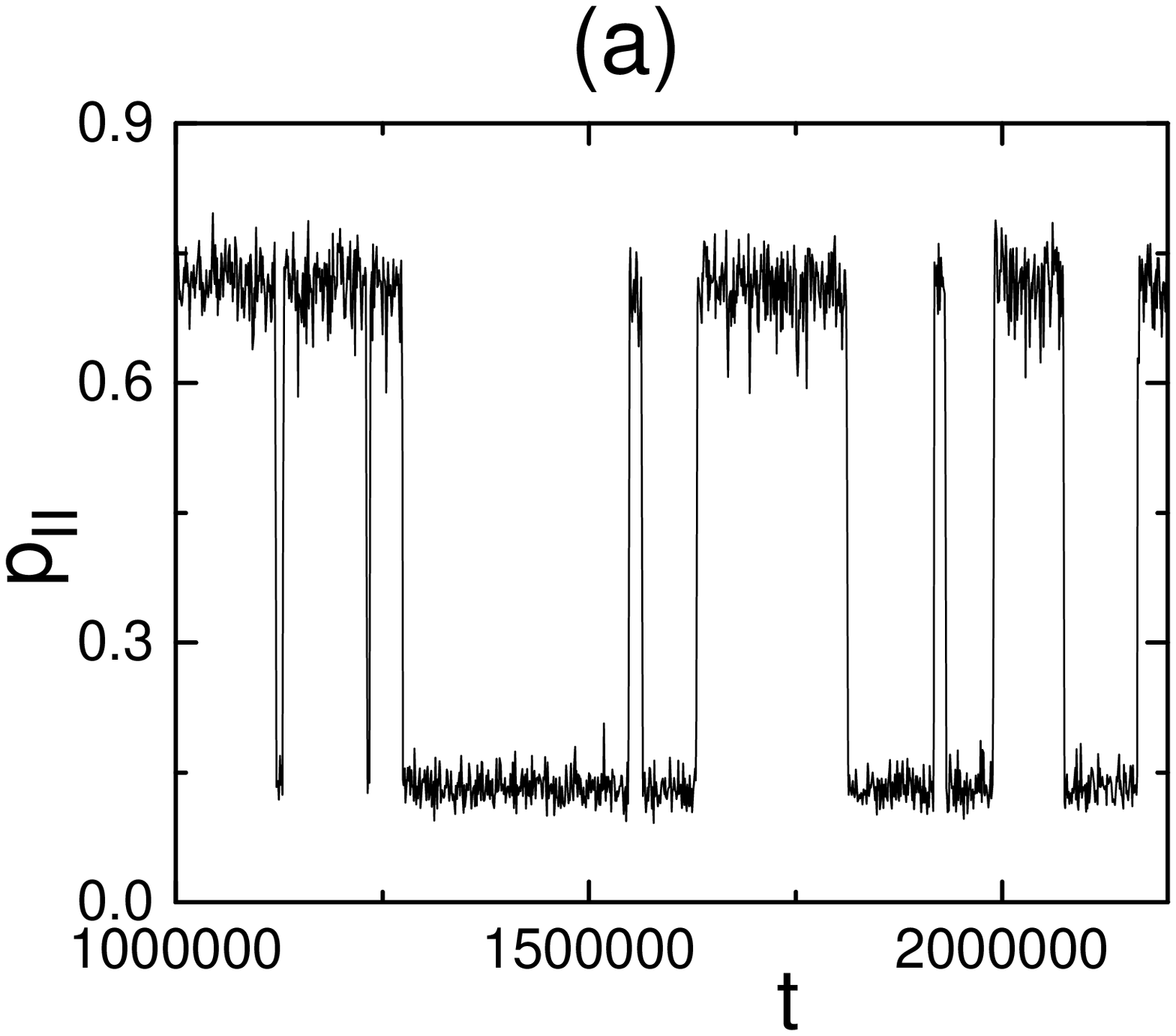}
\includegraphics[width=6.3cm,height=4.3cm,angle=0,clip]{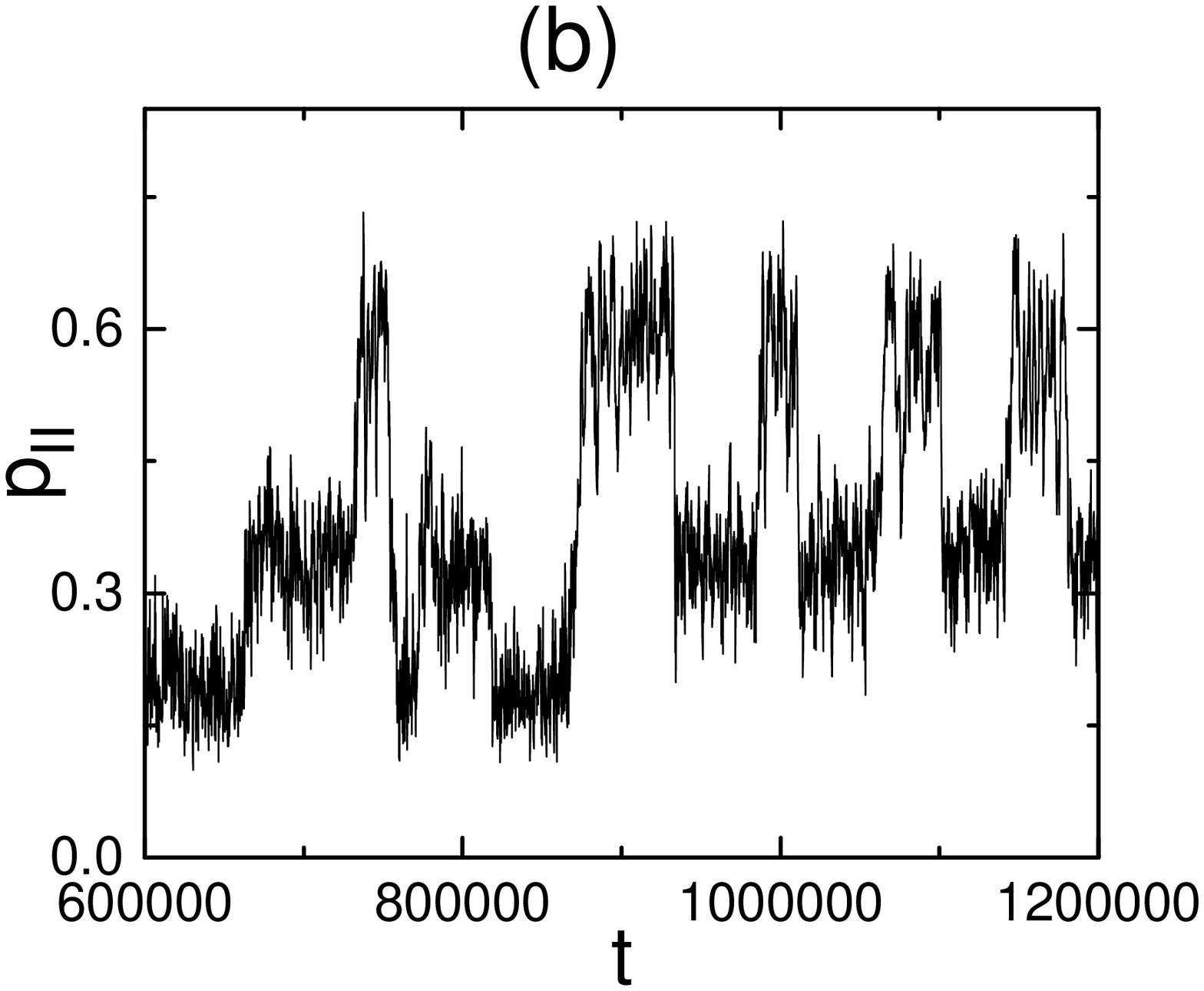}
}\caption{(a) Average density of pluses in middle segment versus
time, inside the  SSB phase ( Panel(a)), and inside the SSB+LDS2
phase (Panel (b)). Panel (a): flips between two quasistable states
$p=1-\beta$ and $p=\beta/2$ are seen. Parameters are
$\alpha=0.45,\beta=0.26,L=20$. Panel (b): Flips between three
quasistable states $p=1-\beta,$ $p=\beta/2$ and $p=1/3$ are seen.
Parameters are
$\alpha=0.45,\beta=0.37,L=100$. }%
\label{Fig_2+3dens}%
\end{figure}

The question of  the accuracy of the MF solution remains.
Discrepancies between the predicted stationary bulk densities and
Monte-Carlo simulations lie within numerical error bars, except for
the LDS2 phase. In this phase the bulk density from Monte Carlo
simulations $p_{II} \approx 0.339$ appears to differ slightly from
the one predicted by mean field $p_{II}=1/3$. This discrepancy
derives from the fact that while the LDS1 and SSB phases are
controlled by uncorrelated inputs  from the outer boundaries, the
LDS2 phase is controlled by the injection rate from the junctions
which could imply correlations in the input.  It appears that such
correlations result in only a small shift of the LDS2
$\leftrightarrow$ LDS1 and LDS2 $\rightarrow$ SSB phase boundaries
to $\alpha \simeq 0.339 < \beta$ and $\beta \simeq 0.339 < \alpha$,
while other phase boundaries remain as in
Fig.\ref{Fig_SSBphasediagram}.  In view of this, we believe that the
mean-field predictions are qualitatively correct and are
quantitatively  very  accurate.

\section{Mechanisms of flips between symmetry-broken states in the SSB phase}
\label{timescale} In a finite system, flips between the two
symmetry-broken states (majority and minority species interchange)
are observed (Fig.\ref{Fig_2+3dens}(a)).  To demonstrate
spontaneous symmetry breaking in the SSB phase and the SSB+LDS2
region Fig.\ref{Fig_SSBphasediagram}, we need to show that the
flipping times between the stable states in finite systems diverge
exponentially with the system size $L$.  For the SSB phase the
mechanism is best illustrated in the SSB region with the
additional condition  $\alpha <1/2$. In this region the plus
profile in section I of Fig.\ref{Fig_AllPhases4}(b) is produced by
a domain wall (a shock) between a region adjacent to the boundary
with density $\alpha$ and a region with density $1-\beta$. The
domain wall is biased to the left and therefore sticks to the left
boundary. (Note that the motion of the plus domain wall does not
affect the state of minus particles.)  In the SSB configuration
the entrance of minuses to the bridge is partially blocked by the
high density $1-\beta$ of pluses. To flip the SSB configuration
Fig.\ref{Fig_AllPhases4}(b), the domain wall of pluses
$p_{I}=1-\beta$ sticking to the left boundary must retreat
(against its bias) back to segment III in order to give an
opportunity for the minuses to take over the key middle segment,
block the entrance of pluses on it, and make the flip.  The time
to wait for such an improbable event grows exponentially with the
system size $L$. Roughly, one can estimate the flipping time as
follows. The shock of pluses $(\alpha,1-\beta)$ is driven to the
right with the rate $r=\beta(1-\beta)/(1-\beta-\alpha)$ and to the
left with the rate $l=\alpha(1-\alpha)/(1-\beta-\alpha)$. In the
SSB region (\ref{SSBstability_region}) $l>r$. Let us represent the
shock position with a phantom particle which hops to the nearest
left/right site with the rates $l$ and $r$. Introducing the
stationary probability $\eta_{k}$ for the phantom particle to be
at site $k$, where $k=0,1,\ldots 2L$ we find
$\eta_{k}l=\eta_{k-1}r$ so that $\eta_{k}=\eta_{0}\left(
r/l\right) ^{k}$. We can estimate the flipping time by
$1/\eta_{2L}$ \cite{MFPT}, giving
\begin{eqnarray*}
t_{flip}^{SSB}  &  \sim \left(  \frac{r}{l}\right) ^{-2L} =
\exp\left(  \kappa L\right)\quad\mbox{where}\quad \kappa
=2\log\frac{\alpha(1-\alpha)} {\beta\left(  1-\beta\right)  }\;.
\end{eqnarray*}
In the rest of the  SSB region (for $\alpha >1/2$) the mechanism is
similar, but
 the density associated with the left hand
boundary is 1/2 rather than $\alpha$. Consequently, in the flipping
time estimate  $\alpha(1-\alpha)$ has to be substituted with $1/4$
for this case.

For the region SSB+LDS2 with three stable phases, flips between the
steady states are also caused by the shock motion, but their
mechanism is different. Firstly, there is no direct transition
between the two SSB states, only through the intermediate
phase SSB1$\Longleftrightarrow$LDS2$\Longleftrightarrow$SSB2, see
Fig.\ref{Fig_2+3dens}(b).
The existence of the intermediate LDS2 state makes the transition much
easier: indeed, to flip from the LDS2 to SSB state, a shock need only
cross one segment as opposed to two segments in direct transitions
between the two symmetry broken states.
This   explains the  big difference in system sizes between
Fig.\ref{Fig_2+3dens}(a) and Fig.\ref{Fig_2+3dens}(b).

\section{Conclusions}

In this work we have introduced a bridge model fed by junctions
where  the input and output streams of the bridge are themselves
TASEPs. This has allowed the phase diagram predicted by the
mean-field approximation to be confirmed by numerical simulations as
being qualitatively correct and rather accurate quantitatively.
The phase diagram exhibits a
number of novel and interesting features that we discussed in
Section II. We note here some further points that will be pursued in
future work

A particularly interesting feature of the phase diagram is that we
have two types of first-order transitions: one with a co-existence line
($\alpha = \beta <1/3$)
and one with a co-existence region (marked LDS2+SSB in
Fig.~\ref{Fig_SSBphasediagram}).  In both cases the co-existence is
between different possible steady states for the systems. The dynamics
of how the system flips between these steady states is rather
intricate but can be understood in terms of domain wall dynamics
(work in progress).

In the phase diagram Fig.~\ref{Fig_SSBphasediagram} a phase resembling
the maximal current phase of the TASEP is absent. We also note that
none of the solutions of the MFE supports maximal current
$j_{+}=j_{-}=1/4$ in the system. The reason is that for the
case $K=1$ considered here, the junctions act as effective
bottlenecks. For larger $K$ (more precisely, for $K>2$), a new phase
reminiscent of the  MC current phase in the TASEP does appear.

\ack This work has been supported by DFG within project KR
1123/1-2, the Albert Einstein Minerva Center for Theoretical
Physics, and the Israel Science Foundation (ISF). The authors
thank The Isaac Newton Institute, Cambridge, where the major part
of this work was done, for hospitality during the {\em Principles
of the Dynamics of Nonequilibrium Systems}
programme. V.P. thanks G. Sch\"utz for discussions.\\


\begin{thebibliography}{99}
\bibitem {MFPT}More accurately,  the flipping time is related to the mean first
passage time of the phantom particle from site $k=0$  to the site
$k=2L$. This will not change the results qualitatively.
\end{thebibliography}
\end{document}